\documentclass[acmlarge,screen,nonacm]{acmart}
\AtBeginDocument{%
  \providecommand\BibTeX{{%
    \normalfont B\kern-0.5em{\scshape i\kern-0.25em b}\kern-0.8em\TeX}}}

\setcopyright{rightsretained}
\copyrightyear{2024}
\acmYear{2024}
\acmDOI{}

\acmConference[CCAI 2024]{CHI 2024 Workshop on Child-centred AI Design}{May 11, 2024}{Honolulu, HI, USA}
\acmBooktitle{CHI 2024 Workshop on Child-centred AI Design, May 11, 2024, Honolulu, HI, USA}

\begin{document}

\title{TaleMate: Exploring the use of Voice Agents for Parent-Child Joint Reading Experiences}

\author{Daniel Vargas-Diaz}
\email{danielvargasdiaz@vt.edu}

\affiliation{%
  \institution{ Computer Science, Virginia Tech}
  \city{Blacksburg}
  \state{VA}
  \country{USA}
}

\author{Jisun Kim  }
\email{jisunk@vt.edu}

\affiliation{%
  \institution{  {Human Development and Family Science, Virginia Tech}}
  \city{Blacksburg}
  \state{VA}
  \country{USA}
}

\author{Sulakna Karunaratna}
\email{sulaknak@vt.edu}

\affiliation{%
  \institution{ Computer Science, Virginia Tech}
  \city{Blacksburg}
  \state{VA}
  \country{USA}
}

\author{Maegan Reinhardt}
\email{maegan@vt.edu}

\affiliation{%
  \institution{  {Human Development and Family Science, Virginia Tech}}
  \city{Blacksburg}
  \state{VA}
  \country{USA}
}

\author{Caroline Hornburg}
\email{chornburg@vt.edu}

\affiliation{%
  \institution{  {Human Development and Family Science, Virginia Tech}}
  \city{Blacksburg}
  \state{VA}
  \country{USA}
}

\author{Koeun Choi}
\email{koeun@vt.edu}

\affiliation{%
  \institution{  {Human Development and Family Science, Virginia Tech}}
  \city{Blacksburg}
  \state{VA}
  \country{USA}
}

\author{Sang Won Lee}
\email{sangwonlee@vt.edu}

\affiliation{%
  \institution{ Computer Science, Virginia Tech}
  \city{Blacksburg}
  \state{VA}
  \country{USA}
}


\begin{abstract}
Joint reading is a key activity for early learners, with caregiver-child interactions such as questioning and feedback playing an essential role in children's cognitive and linguistic development. However, for some parents, actively engaging children in storytelling can be challenging. To address this, we introduce TaleMate—a platform designed to enhance shared reading by leveraging conversational agents that have been shown to support children's engagement and learning. 
TaleMate enables a dynamic, participatory reading experience where parents and children can choose which characters they wish to embody. 
Moreover, the system navigates the challenges posed by digital reading tools, such as decreased parent-child interaction, and builds upon the benefits of traditional and digital reading techniques. TaleMate offers an innovative approach to fostering early reading habits, bridging the gap between traditional joint reading practices and the digital reading landscape.

\end{abstract}



\maketitle

\section{Introduction}

Early development of reading habits significantly influences a child's cognitive and linguistic growth. Shared reading—practices involving a child and a caregiver—is essential in bolstering vocabulary, comprehension, and critical thinking \cite{bus1995joint, Cueva2020, Krmar2014, wesseling2017shared}. Traditional joint reading sessions with printed books often involve the adult reading all dialogues in a single voice and pace, which can be monotonous and cognitively demanding for parents. 
With the advent of digital technologies, the landscape of traditional reading has evolved, bringing new modalities like audiobooks or e-books with voice narration \cite{ewin2021impact}. These digital resources offer numerous benefits, including auditory cues and multimedia effects that can emulate the positive impact of traditional reading \cite{ewin2021impact}.  However, these linear systems do not easily integrate parents into the reading experience, leading to the potential decrease in parent-child interaction observed when digital tools are used. 
Recently, we have witnessed many HCI works that involve conversational agents (CAs) as reading partners as an alternative, offering a new genre of conversation and stimulating interactive questions during reading \cite{Ying2020, Grace2022, Zheng2022, Chang2011}.
To overcome the limitations of linear audiobook reading~\cite{marchetti2018interactivity}, our system enables parents and children to play roles in the book, thereby fostering joint reading and mitigating the disconnection that audiobooks may induce \cite{ewin2021impact}. 

Addressing these challenges and opportunities, we envision an intelligent agent symbiotically collaborating with parents to preserve the benefits of joint reading and make reading a richer experience. This paper presents, TaleMate, a system designed to facilitate a rich reading environment. We introduce Conversational Agents (CAs) into the reading process, reading a book collaboratively with parents and potentially with children like a theatrical play and adding a novel selection process of choosing an actor for each character in the book, thereby enhancing the interactive experience. Our system features an audiobook mode, offering a unique sensory experience and minimizing device-related distractions. We present the current design of our system and our ongoing efforts to evaluate the user experience of our platform and the potential benefits it could offer when compared to other reading experiences.

\section{Design of the system}

The design of TaleMate includes two key components of the system:
\begin{itemize}
\item Voice agents integration and user selection to enrich and personalize the reading experience.
\item Employing books that enable the user to participate actively, thereby alternating turns in the reading process.
\end{itemize}


\subsection{Integration of Voice Agent and Assignment Process}

TaleMate offers an enriched reading experience through the integration of voice agents. The system houses six distinct voice agents, each boasting a unique artificial voice. Within the system, these agents are referred to as "Mates", each represented by an animated sprite mimicking a book, which has been shown to boost engagement and affinity \cite{Grace2022}. The role of each Mate is to facilitate parents and children in assigning a particular voice to each character in the book. The multi-voice feature draws on research that suggests the integration of audio cues and multimedia effects can uplift reading comprehension \cite{BUS201579}. We provide six unique voices that can be designated to each character in the ebook. As illustrated in Figure \ref{fig:character_selection}, both the adult and child (number 7) can actively participate in the voice agent assignment process. This design allows users to constructively design their reading experience by assigning voices to each character. The avatars representing the adult and child are designed using the Avataaars-generator library \cite{Fang2021}, promoting inclusive avatar generation.

Figure \ref{fig:character_selection} presents the screen where users can assign voices in our system. Users can drag and drop each reader (\ref{fig:character_selection}-4) and assign them to each character (\ref{fig:character_selection}-6) by drag-and-dropping a reader from the left panel into these boxes to assign roles to the respective characters. To further aid the user, instructions are also presented at the top of the page (\ref{fig:character_selection}-2). Users can click the "Play" icon (\ref{fig:character_selection}-5) below the picture to hear the sound of an agent saying, "Hello, I am Mate X," where X denotes the Mate's number. This feature lets users sample each Mate's voice before making a selection. These voices are generated using Google's Speech to Text API. Upon completion, users can click the button at the top-right corner (\ref{fig:character_selection}-3) to proceed to the story or one on the top-left corner (\ref{fig:character_selection}-1) for going back to the home screen. The system is designed such that voice agents, adults, and children alternate in reading specific character dialogues.

%

\subsection{Promoting Joint Media Reading}
\begin{figure}[t]
    \centering
    \begin{minipage}{0.48\linewidth}
        \includegraphics[width=\linewidth]{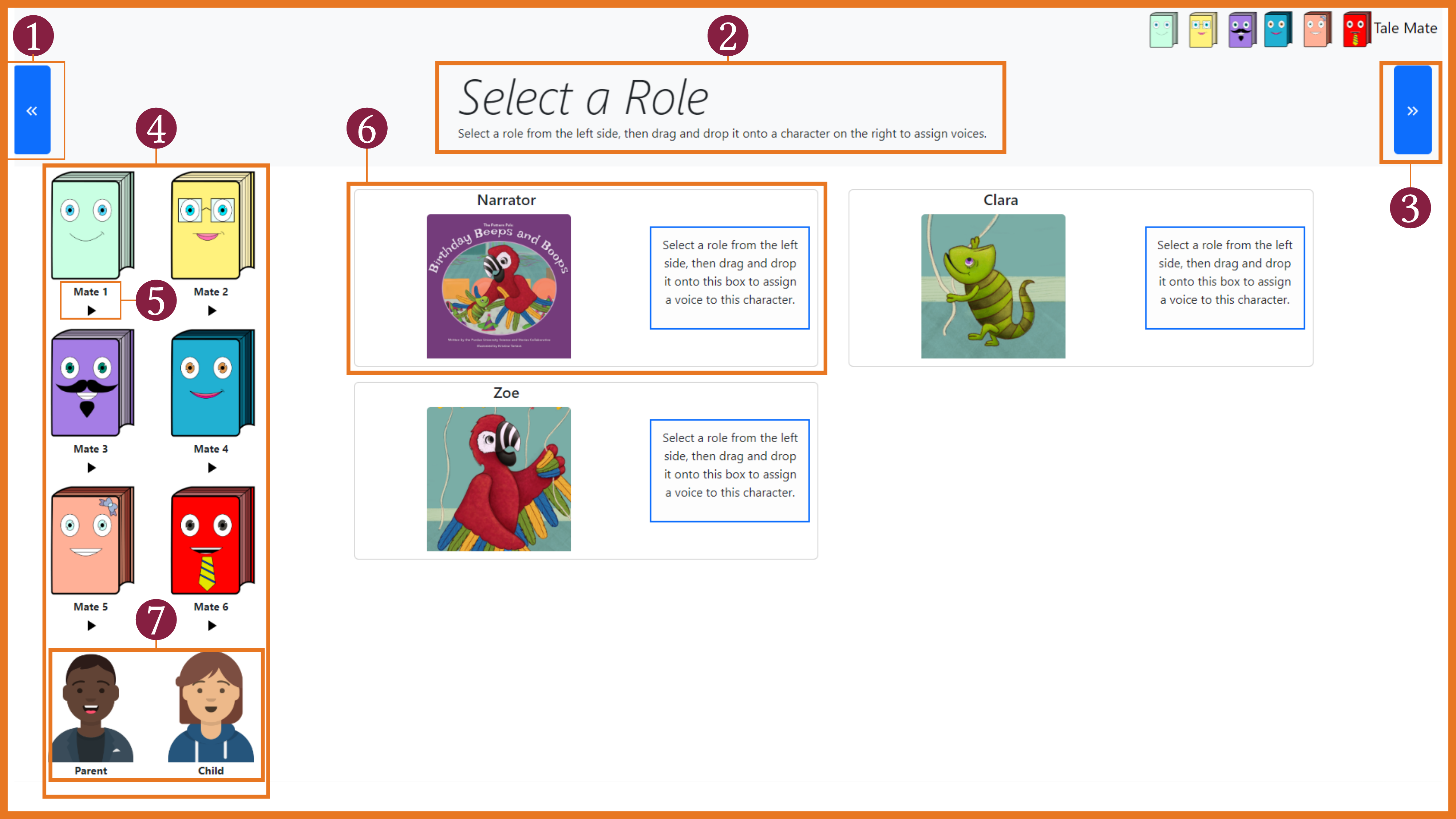}
        \caption{Screen facilitating the assignment of roles (4) to book characters (6)}
        \Description{The screen supports the selection of characters and roles.}
        \label{fig:character_selection}
    \end{minipage}
    \hfill 
    \begin{minipage}{0.48\linewidth}
        \includegraphics[width=\linewidth]{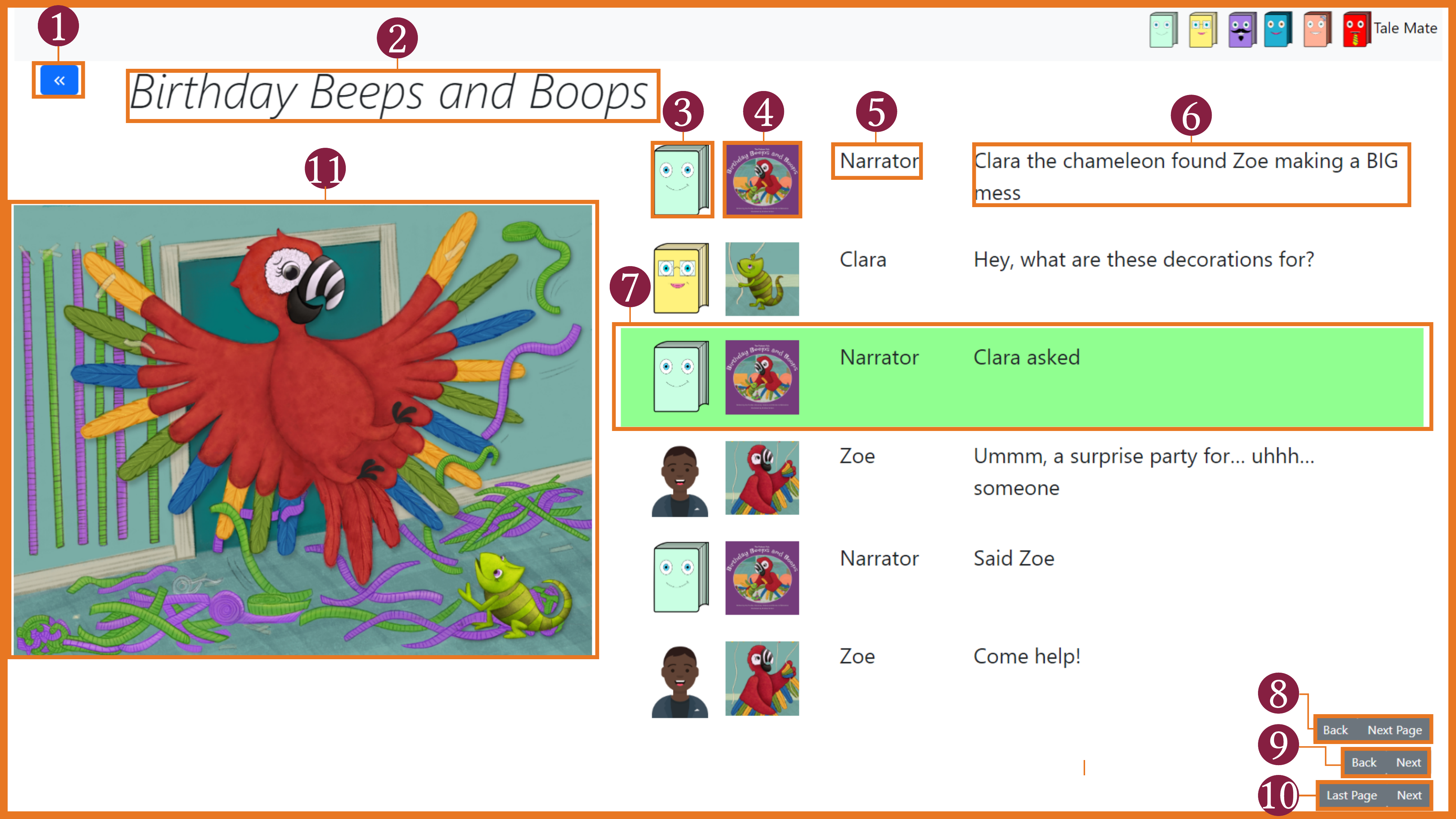}
        \caption{Reading screen where the user and voice agents alternate turns to read the story}
        \Description{Screen where users and voice agents interact.}
        \label{fig:reading_screen}
    \end{minipage}
\end{figure}

In TaleMate, parents pace reading and determine the timing of voice agents' reading each line. Thus, they can seamlessly integrate parent-child interaction, such as questions and answers or non-diegetic conversation into reading. If a parent presses the "Next" button (\ref{fig:reading_screen}-9), the system will use the corresponding agent to read the particular line. If it is one of the parent's or children's turn, it will highlight the line and wait for them to read the line. 

To encourage collaborative joint media reading, we integrated the book "Birthday Beeps and Boops" from the "The Pattern Pals" series, published by the Purdue Science and Stories Collaborative \cite{wijns2022stimulating}. We selected this book for its age-appropriate content for our target audience (Ages 3-6) and the distinct dialogue for each character, as opposed there being only a narrator. A randomized controlled study has shown that the selected books are effective in improving children's patterning ability \cite{wijns2022stimulating}. 

To facilitate revisiting passages, the system shows different options (one of \ref{fig:reading_screen}-8, 9, and 10) depending on their reading progress, simplifying the parents' interaction while pacing reading. Other elements on the reading screen include the book's title (\ref{fig:reading_screen}-2) and a button to return to the home screen (\ref{fig:reading_screen}-1). To display the book's content, we have a line for each character (\ref{fig:reading_screen}-7), with the currently reading character's or agent's line highlighted in green. Each line includes the role responsible for reading it (\ref{fig:reading_screen}-3), the image of the character reading that line (\ref{fig:reading_screen}-4), the character's name (\ref{fig:reading_screen}-5), and the character's dialogue (\ref{fig:reading_screen}--6). 
The system is designed so that users and voice agents take turns reading dialogues in the book, based on the character assignments made on the character selection screen (Figure \ref{fig:character_selection}). For example, when the user clicks the next button, "Mate 1" gets the first turn and reads the Narrator's dialogue. Once "Mate 1" finishes, the user can click next again, and it will be "Mate 2"'s turn, who plays the role of Clara. The cycle continues with each click of the next button, providing the turn to the respective characters. Depending on when the parent presses the button, parents can have a conversation about the book freely. 

\section{Ongoing efforts}
We are currently conducting the initial user studies for our system, focusing on evaluating the shared user experience of parents and children. Our experiment consists of two reading sessions. In the first session, the parent reads the book with our voice agents. In the second session, children have the opportunity to portray a character from the book under the guidance of their parents by either reading or repeating the characters' dialogues, depending on their reading ability. Following each session, we conduct an interview with both the parent and child to assess their experience with the system. These interviews aim to understand the usability aspects of the system, such as ease of use, perceived value of the platform, and overall satisfaction. Potential questions include, "How easy was it to use the platform?" and "Did the platform enhance your reading experience?" The study primarily employs a qualitative approach, and the collected responses are analyzed using thematic analysis techniques to discern patterns, user perceptions of the platform, and potential areas for improvement.


\bibliographystyle{ACM-Reference-Format}
\bibliography{references}

\appendix

\end{document}